\documentstyle[aps,preprint,epsf]{revtex}

\begin{document}
\draft \tightenlines

\def\be{\begin{equation}}
\def\ee{\end{equation}}
\def\bea{\begin{eqnarray}}
\def\eea{\end{eqnarray}}
\def\pd{\partial}
\def\a{\alpha}
\def\b{\beta}
\def\g{\gamma}
\def\d{\delta}
\def\m{\mu}
\def\n{\nu}
\def\t{\tau}
\def\l{\lambda}

\def\s{\sigma}
\def\e{\epsilon}
\def\scri{\mathcal{J}}
\def\cM{\mathcal{M}}
\def\tcM{\tilde{\mathcal{M}}}
\def\RR{\mathbb{R}}

\preprint{KUNSAN-TP-00-4}

\title{Dilaton-driven brane inflation in type IIB string theory}

\author{ Jin Young Kim\footnote{Electronic address:
jykim@ks.kunsan.ac.kr}}
\address{Department of Physics, Kunsan National University,
Kunsan, Chonbuk 573-701, Korea}
\maketitle

\begin{abstract}

We consider the cosmological evolution of the three-brane in the
background of type IIB string theory. For two different
backgrounds which give nontrivial dilaton profile we have derived
the Friedman-like equations. These give the cosmological evolution
which is similar to the one by matter density on the universe
brane. The effective density blows up as we move towards the
singularity showing the initial singularity problem. The analysis
shows that when there is axion field in the ambient space the
recollapsing of the universe occurs faster compared with the case
without axion field.

\end{abstract}

\pacs{PACS number(s): 11.25.-w, 11.10.Kk, 98.50.Cq }

\newpage

\section{Introduction}

The idea that our universe might be a domain wall embedded in a
higher dimensional space \cite{RT} has attracted much interest
recently. It is possible that the fundamental scale of gravity can
be lowered to the electroweak scale by introducing large extra
dimensions \cite{HDD}. Randall and Sundrum \cite{RS} proposed
scenarios that our observed universe is embedded in a
five-dimensional bulk, in which the background metric is curved
along the extra dimension due to the negative bulk cosmological
constant. These models have been studied extensively because they
might provide the solution to the gauge hierarchy problem and
cosmological constant problem \cite{KVK}. The cosmology can be
different from the conventional four dimensional one.

One can naturally attempt to justify the scenario within a well
defined framework for higher dimensional quantum theory of gravity
such as string theory. Many attempts have been made to apply this
idea to string theory in the context with D-branes \cite{Pol},
where the standard model gauge bosons as well as charged matter
arise as fluctuations of the D-branes. An early example of this is
the Horava-Witten picture for the nonperturbative heterotic $E_8
\times E_8$ string \cite{HW}. The spacetime includes a compact
dimension with an orbifold structure.  Matter is confined to the
hypersurface which forms the boundaries of the spacetime. Within
the string theory context, it is natural that our observable
four-dimensional world is a three-brane embedded in ten
dimensional string. In such theories, one of the important issue
is the cosmological evolution of our universe. Many cosmological
models associated to brane universe have been suggested.  The
models can be classified into two categories. The first is that
the domain walls (branes) are static solution of the underlying
theory and the cosmological evolution of our universe is due the
time evolution of energy density on the domain wall (brane)
\cite{Tye}. The second is that the cosmological evolution of our
universe is due to the motion of our brane-world in the background
of gravitational field of the bulk \cite{CR,Kiritsis}. We will
focus on the second approach in this paper.

It is shown that the motion of the brane in ambient space induces
cosmological expansion (or contraction) on our universe simulating
various kinds of matter or a cosmological constant. In other words
the cosmological expansion is not due to energy density on our
universe but somewhere else. This is the idea by mirage cosmology
\cite{Kiritsis}. Friedman-like equations were derived for various
bulk background field solutions. In \cite{PP}, the motion of a
three-brane, in a background of type 0 theory was examined.

In this paper, employing the formalism of \cite{Kiritsis}, we will
study how the presence of matter field on the background geometry
affects the cosmological evolution of the brane universe. More
specifically we will consider cosmological evolution of type IIB
theory with two different background geometries, one without axion
and the other with axion.  We will compare the two and see the
difference.

The organization of the paper is as follows. In Sec. II we will
briefly review the formalism of mirage cosmology in ref.
\cite{Kiritsis} and set up some preliminaries for our calculation.
In Sec. III we consider the type IIB theory and its background
solution with and without axion field. In Sec. IV, using the
background solutions of Sec. III, we find the cosmological
evolution of the three-brane under the background. Finally section
V is devoted to conclusions and discussion.

\section{Formalism }
In this section, we consider a probe brane moving in a generic
static spherically symmetric background. We ignore its back
reaction to the ambient space.  As the brane moves in a geodesic,
the induced world-volume mertic becomes a function of time. The
cosmological evolution is possible from the brane resident point
of view. We will focus on a D3-brane case. For this purpose we
parametrize the metric of a D3-brane as
 \be
 ds^2_{10} = g_{00} (r) dt^2 + g(r) (d \vec x)^2 + g_{rr} (r) dr^2
 + g_S(r) d\Omega^2_5 ,
 \label{ds10gen}
 \ee
and there are dilaton field $\phi$ as well as RR (Ramond-Ramond)
background $C(r) = C_{0 \dots 3} (r)$.
 The probe brane will in general move in this background and its
dynamics is governed by the Dirac-Born-Infeld (DBI) action. In
maximally supersymmetric case, ignoring the fermions, it is given
by
 \bea
 S & = & T_3 \int d^4 \xi e^{- \phi} \sqrt{-det( {\hat
G}_{\alpha\beta} + (2 \pi \alpha^\prime) F_{\alpha\beta} -
B_{\alpha\beta} ) }          \nonumber \\
 & + & T_3 \int d^4 \xi {\hat C}_4 + {\rm anomaly ~terms} ,
 \label{DBI}
 \eea
 where the induced metric on the brane is
 \be
 \hat G_{\alpha\beta} = G_{\mu\nu}
 { {\partial x^\mu} \over {\partial \xi^\alpha} }
 { {\partial x^\nu} \over {\partial \xi^\beta} }
 \ee
with similar expressions for other fields. Generally the motion of
a probe D3-brane have a nonzero angular momentum in the transverse
directions. We can write the relevant part of the Lagrangian, in
the static gauge $x^\alpha = \xi^\alpha ~ ( \alpha = 0,1,2,3)$, as
 \be
 L = \sqrt{ A(r) - B(r) {\dot r}^2 - D(r) h_{ij} {\dot \varphi}^i
 {\dot \varphi}^j } - C(r),
 \ee
where $ h_{ij} \varphi^i \varphi^j $ is the line element on the
unit five-sphere ($i,j=5, \cdots, 9$),
 \be
 A(r)=g^3(r)|g_{00}(r)|e^{-2\phi} , ~~
 B(r)=g^3(r)g_{rr}(r)e^{-2\phi} , ~~
 D(r)=g^3(r)g_S(r)e^{-2\phi} ,
 \label{ABD}
 \ee
and $C(r)$ is the RR background. The momenta of the system are
given by
 \bea
 && p_r = -{B(r) \dot r \over
 \sqrt{A(r)-B(r)\dot r^2 - D(r) h_{ij} {\dot \varphi}^i
 {\dot \varphi}^j } } ,
    \nonumber \\
 && p_i  = -{D(r) h_{ij} \dot \varphi^j \over
 \sqrt{A(r)-B(r)\dot r^2 - D(r) h_{ij} {\dot \varphi}^i
 {\dot \varphi}^j } } .
 \eea
Calculating the Hamiltonian and demanding the conservation of
energy, we have
 \be
 H = C(r) - { A(r) \over \sqrt{A(r)-B(r)\dot r ^2-D(r)h_{ij} {\dot
 \varphi}^i {\dot \varphi}^j} } = - E ,
 \label{Hamil}
 \ee
 where $E$ is the total energy of the brane.
 Also from the conservation of the total angular momentum $h^{ij}p_ip_j=\ell^2$,
 we have
 \be
 h_{ij} {\dot \varphi}^i {\dot \varphi}^j = {\ell^2(A(r)-B(r)\dot
 r^2) \over  D(r)(D(r)+\ell^2)} .
 \label{lcons}
 \ee
Substituting equation (\ref{lcons}) into (\ref{Hamil}) and solving
with respect to $\dot r^2$, we have the equation for the radial
variable as
 \be
 {\dot r}^2 = {A\over B} \left\{
 1-{A\over (C+E)^2}{D+\ell^2 \over D}
 \right\} .
 \label{rsol}
 \ee
 Plugging equation (\ref{rsol}) back into (\ref{lcons}), we
 have the equation for the angular variable
 \be
 h_{ij} {\dot \varphi}^i {\dot \varphi}^j
 = {A^2\ell^2\over D^2(C+E)^2} .
 \label{lsol}
 \ee

The induced four-dimensional metric on the three-brane universe is
 \be
 d \hat s^2_{4d} = (g_{00}+g_{rr} \dot r^2
 + g_S h_{ij} {\dot \varphi}^i {\dot \varphi}^j )
 dt^2+g(d\vec x)^2 .
 \label{4dmet}
 \ee
Using equation (\ref{rsol}) and (\ref{lsol}), this reduces to
 \be
 d\hat s^2_{4d}=-{g^2_{00}g^3 e^{-2\phi} \over (C+E)^2}dt^2
 + g(d\vec x)^2 = -d \eta^2 +g(r(\eta))(d\vec x)^2 ,
 \ee
 where we defined, for the standard form of
a flat expanding universe, the cosmic time $\eta$ as
 \be
 d\eta={|g_{00}|g^{3/2}e^{-\phi}\over |C+E|}dt .
 \ee
If we define the scale factor as $a^2=g$, we can calculate, from
the analogue of the four-dimensional Friedman equation, the Hubble
constant $H={\dot a / a}$
 \be
 \left({\dot a \over a}\right)^2
 ={(C+E)^2g_Se^{2\phi}-|g_{00}|(g_Sg^3 + \ell^2e^{2\phi})
 \over 4|g_{00}|g_{rr}g_Sg^3 }
 \left({g'\over g}\right)^2 ,
 \label{hub}
 \ee
 where the dot denotes the derivative
with respect to cosmic time and the prime denotes the derivative
with respect to $r$. The right hand side of (\ref{hub}) can be
interpreted as the effective matter density on the probe brane
 \be
 {8\pi \over 3}\rho_{\rm eff}
 ={(C+E)^2g_Se^{2\phi}-|g_{00}|(g_Sg^3 + \ell^2e^{2\phi})
 \over 4|g_{00}|g_{rr}g_Sg^3 }
 \left({g'\over g}\right)^2 .
 \label{hubble}
 \ee
We have also
 \bea {\ddot a\over a}
 &=& \left(1+{g\over g'}{\partial\over \partial r}\right)
 {(C+E)^2g_Se^{2\phi}-|g_{00}|(g_Sg^3+ \ell^2e^{2\phi})
   \over 4|g_{00}|g_{rr}g_Sg^3}
  \left({g'\over g}\right)^2 \nonumber \\
  &=& \left( 1+{1 \over 2} a {\partial\over \partial a} \right)
  {8\pi \over 3}\rho_{\rm eff} .
 \eea
Equating the above to
 $-(4\pi / 3)(\rho_{\rm eff} + 3p_{\rm eff})$,
 we can find the effective pressure
 \be
 p_{\rm eff} = - \rho_{\rm eff}
 - {1 \over 3}  a {\partial\over \partial a} \rho_{\rm eff}.
 \ee
The apparent scalar curvature of the four-dimensional universe is
 \be
 R_{4d}=6 \left\{ {\ddot a\over a}
 +  \left({\dot a \over a}\right)^2 \right\}
 = 8\pi\left(4+a\partial_a\right)\rho_{\rm eff} .
 \label{4dcurv}
 \ee
We have given the formalism for simple D3-brane case. The geodesic
motion of Dp-brane in the background of the ${\rm Dp}'$-brane with
$p' > p$ can be generalized easily. In the case $p= p'$, there
exists the additional Wess-Zumino term $T_p \int {\hat C}_{p+1}$
in the DBI action which modifies the equation of the probe brane
as well as the induced metric. This modification turn out to be
the shift $E \to E + C$ where $C = C_{0 \dots p}$ \cite{Kiritsis}.

\section{The type IIB background solution}

Here we will consider the background geometry of type IIB theory
with five-form flux through an $S^5$. We will also assume, for the
metric, (3+1)-dimensional Poincar\'e invariance $ISO(1,3)$ since
we need the theory defined on the Minkowski space-time. In
addition we will preserve the $SO(6)$ symmetry of the $AdS_5
\times S^5$. As a result, the $ISO(1,3) \times SO(6)$ invariant
ten-dimensional metric with $N$ units of five-form flux through an
$S^5$, in the Einstein frame, can be written as
 \bea
   ds_{10}^2 &=& \hat{g}_{MN} dx^M dx^N
    = e^{-{10 \over 3} \chi + 2\sigma} (-dt^2 + d {\vec x}^2 + dr^2)
       + L^2 e^{2\chi} d\Omega_5^2 , \\
   F_5 &=& {N \sqrt{\pi} \over 2 {\rm Vol} S^5}
     \left( {\rm vol}_{S^5} + * {\rm vol}_{S^5} \right) ,
      \label{fiveform}
 \eea
 where $\chi$, $\sigma$, and also the dilaton $\phi$ and the axion $\eta$ are
allowed to depend only on the radial coordinate $r$.

 The equations of motion in type IIB supergravity, truncated to the fields
 of our interests, are given by \cite{GSW}
  \bea
   \hat{\nabla}^2 \phi & = & - e^{2\phi} (\partial \eta)^2 , \nonumber \\
   \hat{\nabla}^2 \eta & = & -2 (\partial_M \phi) (\partial^M \eta) , \\
   \hat{R}_{MN} & = & {1 \over 2} \partial_M \phi \partial_N \phi
    - {1 \over 2} e^{2\phi} \partial_M \eta \partial_N \eta +
    {\kappa^2 \over 6} F_{MKLPQ} F_N{}^{KLPQ} , \nonumber
  \eea
where hat means that the operators are expressed in
ten-dimensional terms and $M,N, \cdots = 0, \cdots , 9$. The
equation of motion for the five-form field is
 \be
 {\hat \nabla}_M  F^{MKLPQ} = 0 ,
 \label{RReq}
 \ee
which is satisfied with the self-duality condition
(\ref{fiveform}). The Einstein equation in $S^5$ direction is
  \be
   {\hat R}_{ij} =\left( {\kappa N \over 2 \pi^{5/2}} \right)^2 g_{ij}, ~~~~~
   i,j= 5, \cdots ,9
  \ee
which is automatically satisfied if
  \be
   L^4 = {\kappa N \over 2 \pi^{5/2}} .
  \ee
 The remaining equations can be expressed in purely five-dimensional terms
  \bea
  \nabla^2 \phi &=& - e^{2\phi} (\partial_\mu \eta)^2 , \nonumber \\
  \nabla^2 \eta &=& - 2 (\partial_\mu \phi) (\partial^\mu \eta) , \nonumber \\
  \nabla^2 \chi &=& {4 \over L^2} \left( e^{-{16 \over 3} \chi} -
    e^{-{40 \over 3} \chi} \right) , \label{5deqns}\\
   R_{\mu\nu} &=& {1 \over 2} \partial_\mu \phi \partial_\nu \phi
     - {1 \over 2} e^{2\phi} \partial_\mu \eta \partial_\nu \eta +
    {40 \over 3} \partial_\mu \chi \partial_\nu \chi -
    {g_{\mu\nu} \over L^2} \left( {20 \over 3} e^{-{16 \over 3} \chi} -
     {8 \over 3} e^{-{40 \over 3} \chi} \right) , \nonumber
  \eea
 where the new metric
  \be
   ds_5^2 = g_{\mu\nu} dx^\mu dx^\nu =
    e^{2\sigma} (-dt^2 + d{\vec x}^2 + dr^2)
   \label{fivemet}
  \ee
 should be used to compute $R_{\mu\nu}$ and to contract indices.  The
 equations in (\ref{5deqns}) can be derived from the
 five-dimensional action \cite{GKT}
  \be
   S = {1 \over 2 \kappa_5^2}
    \int d^5 x \sqrt{-g} \left\{ R - {1 \over 2} (\nabla \phi)^2 +
     {1 \over 2} e^{2\phi} (\nabla \eta)^2 -
     {40 \over 3} (\nabla \chi)^2 + {1 \over L^2}
      \left( 20 e^{-{16\over 3} \chi} - 8 e^{-{40\over 3} \chi} \right)
      \right\} ,
  \ee
 where the gravitational couplings $\kappa_5$ and $\kappa$ in five and ten
dimensions are related by
  \be
   {1 \over 2 \kappa_5^2} = {\pi^3 L^5 \over 2 \kappa^2} =
    {N^2 \over 8 \pi^2} {1 \over L^3} .
  \ee
Note the minus sign in front of the axion kinetic term, which is
the result of the Hodge-duality rotation of the type-IIB nine-form
\cite{GGP}.

\subsection{Solution without axion}

In general one can reduce the equations of motion (\ref{5deqns})
 to a set of coupled non-linear second order ordinary
differential equations in $\phi$, $\eta$, $\chi$, and $\sigma$.
These equations are too complicated to solve in general, but there
is an obvious simplification with $\chi=\eta=0$ \cite{Gub}.  Then
the equations in (\ref{5deqns}) become much simpler

  \bea
  && e^{-5 \sigma} \partial_r e^{3 \sigma} \partial_r \phi = 0,
                        \label{dilwo}  \\
  &&  \partial_r^2 \sigma + 3 (\partial_r \sigma)^2
    = {4 \over L^2} e^{2\sigma},  \label{Rttwo}  \\
  && 4 \partial_r^2 \sigma =
  - {1 \over 2} (\partial_r \phi)^2 + {4 \over L^2} e^{2\sigma}.
            \label{Rrrwo}
  \eea
Equations (\ref{Rttwo}) and (\ref{Rrrwo}) are obtained from $(tt)$
and $(rr)$ components of the Einstein equation. Because $\chi = 0$
there is no distinction between the five-dimensional Einstein
metric and ten-dimensional metric restricted to the
five-dimensional noncompact subspace. The equation (\ref{dilwo})
can be integrated to give
 \be
 \phi (r) = \phi_\infty + { B \over L}
   \int_0^r d \tilde r e^{-3 \sigma (\tilde r) } ,
 \label{dilwom}
 \ee
where $\phi_\infty$ is the value of the dilaton at the boundary of
the asymptotically $AdS_5$ geometry and $B$ is an integration
constant. Substituting (\ref{dilwom}) into (\ref{Rttwo})
 and (\ref{Rrrwo}), defining new variable $u \equiv r/L$, we obtain

 \bea
  &&   (\partial_u \sigma)^2
    = e^{2\sigma} + {B^2 \over 24} e^{-6 \sigma} \label{Rttwom} , \\
  && \partial_u^2 \sigma = e^{2\sigma} - {B^2 \over 8} e^{-6 \sigma} .
            \label{Rrrwom}
 \eea
 The second equation follows from differentiating the first, so we see that
(\ref{dilwom}), (\ref{Rttwom}) and (\ref{Rrrwom}) are
 consistent system of equations despite being overdetermined.

One can understand (\ref{Rttwom}) as a mechanical analog of a
classical particle with unit mass moving in the potential
 \be
 V(\sigma) = - {1 \over 2}
 e^{2\sigma} - {B^2 \over 48} e^{-6\sigma},
 \ee
 with zero enegry. If $B=0$, the solution is pure $AdS_5$ with constant dilaton.
 To have a solution with nonconstant dilaton, we take $B > 0$.
 However, the $B\neq 0$ geometry is geodesically incomplete and singular
at some point $u=u_0$.  To find $u_0$ explicitly, we integrate
equation (\ref{Rttwom})
  \be
   u =  \int_\sigma^\infty {d\tilde\sigma \over
        \sqrt{e^{2\tilde\sigma} + {B^2 \over 24} e^{-6\tilde\sigma}}}
     = {3^{1/8} \Gamma(3/8) \Gamma(1/8) \over 8^{7/8} \sqrt{\pi} B^{1/4}} -
        \sqrt{{8 \over 3}} {e^{3\sigma} \over B}
        F\left( {3 \over 8},{1 \over 2};{11 \over 8};
         -{24 e^{8 \sigma} \over B^2} \right) ,
   \label{usol}
  \ee
 where $F(\alpha,\beta;\gamma;z)$ is the usual hypergeometric function.
The second term vanishes as $\sigma \to -\infty$, so we find
 \be
 u_0 = {3^{1/8} \Gamma(3/8) \Gamma(1/8) \over 8^{7/8} \sqrt{\pi}
 B^{1/4}}.        \label{u0sol}
 \ee
Also we find the dilaton in terms of $\sigma$ by solving the
equation (\ref{dilwom})
 \bea
 \phi && = \phi_\infty + \sqrt{3 \over 2}
 \coth^{-1} \sqrt{ 1 + (24/B^2)e^{8 \sigma} }   \nonumber  \\
     && = \phi_\infty + {1 \over 2} \sqrt{3 \over 2}
   \ln {       \sqrt{ 1 + (24/B^2) e^{8 \sigma} } + 1
           \over \sqrt{ 1 + (24/B^2) e^{8 \sigma} } - 1    } .
  \label{dilwosol}
 \eea
We can write the ten-dimensional Einstein metric explicitly if we
use $\sigma$ as the radial variable

  \be
   ds_{10}^2 = e^{2\sigma} (-dt^2 + d{\vec x}^2 ) +
    {L^2 d\sigma^2 \over 1 + (B^2 / 24) e^{-8\sigma}} +
    L^2 d\Omega_5^2 \ .
   \label{5dmetwo}
  \ee
 We can cancel the factors of ${B^2 / 24}$ from
 (\ref{dilwosol}) and (\ref{5dmetwo}) by replacing
by $\sigma \to \sigma + (1 / 8) \ln (B^2 / 24)$,
 $t \to (B^2/24)^{-1/8} t$ and $x_i \to (B^2/24)^{-1/8} x_i$ for
 $i = 1,2,3$. If we also
rescale $r \to ( B^2 / 24)^{-1/8} r$ then the net result is the
same as if we had set $B^2 = 24$. Choice of the radial coordinate
in $AdS_5$ corresponds to choice of one out of a given class of
conformally equivalent boundary metrics. Thus the freedom to
change $B$ in the solution corresponds to the asymptotic scale
invariance of the boundary theory.

\subsection{Solution with axion}

In the previous subsection we described the simplest case in which
one can have a solution with nontrivial dilaton. Here we will
consider the case with only $\chi=0$ to find the solution with
nontrivial axion field \cite{Kesf}. The equations in
(\ref{5deqns}) can be written
 \bea
 && {1 \over \sqrt{-g} } \partial_\mu \left(\sqrt{-g}
 g^{\mu\nu} \partial_\nu \phi\right)
 = - e^{2\phi} \partial_\nu \eta \partial_\mu \eta g^{\mu\nu} ,
 \label{5ddil}  \\
 && {1\over \sqrt{-g} } \partial_\mu \left(\sqrt{-g} g^{\mu\nu} e^{2\phi}
 \partial_\nu \eta\right)
 =0 ,  \label{5daxi}          \\
 &&R_{\mu\nu} = - {4 \over L^2} g_{\mu\nu}
   + {1 \over 2} \partial_\mu \phi \partial_\nu \phi
 - {1 \over 2} e^{2\phi} \partial_\mu \eta \partial_\nu \eta .
 \label{5dRmunu}
 \eea
Integral of the axion in equation (\ref{5daxi}) is
 \be
 \eta' = \eta_0  e^{-3 \sigma}e^{-2\phi} ,
 \label{axifirst}
 \ee
where the prime denotes derivative with respect to
 $u=r / L$ and $\eta_0$ is an integration constant.
Inserting this expression for $\eta$ into equation (\ref{5ddil})
we obtain the differential equation for dilaton
 \be
 e^{3 \sigma} ( e^{3 \sigma} \phi')'
 = - \eta_0^2 e^{-2\phi},  \label{dil2}
 \ee
 and integrating once we have
 \be
  e^{6 \sigma} \phi'^2
 =  \eta_0^2 e^{-2\phi} + {\tilde B}^2,  \label{dilfirst}
 \ee
 where ${\tilde B}$ is another arbitrary constant.
Equations (\ref{axifirst}) and (\ref{dilfirst}) are sufficient to
proceed and solve for the function $\sigma(u)$ that appears in the
metric (\ref{5dRmunu}). The Einstein equation (\ref{5dRmunu})
becomes
  \bea
    \partial_u^2 \sigma &&=  e^{2\sigma}
   - {1 \over 8} (\partial_u \phi)^2
   + {1 \over 8} (\partial_u \eta)^2   \label{Ruu}  \\
    \partial_u^2 \sigma + 3 (\partial_u \sigma)^2 &&
    = 4 e^{2\sigma}.  \label{Rtt}
  \eea
 Inserting (\ref{axifirst}) and (\ref{dilfirst}) into (\ref{Ruu})
 and (\ref{Rtt}), we obtain
 \bea
  \partial_u^2 \sigma &&=  e^{2\sigma}
   - {{\tilde B}^2 \over 8} e^{-6 \sigma}, \label{Ruude}   \\
  (\partial_u \sigma)^2 &&
    =  e^{2\sigma} - {{\tilde B}^2 \over 24} e^{-6 \sigma}.  \label{Rttde}
 \eea
The above equations are exactly the same as those without axion
(see (\ref{Rttwom}) and (\ref{Rrrwom}) ). So the expressions  for
$u$ and $u_0$ in (\ref{usol}) and (\ref{u0sol}) are still valid in
the presence of axion field. Also we find the dilaton in terms of
$\sigma$ by solving the equation (\ref{dilfirst})
 \bea
 e^\phi && = { {\tilde B} \over  \eta_0} \sinh \left\{
    \ln \left( {       \sqrt{ 1 + (24/{\tilde B}^2) e^{8 \sigma} } + 1
            \over \sqrt{ 1 + (24/{\tilde B}^2) e^{8 \sigma} } - 1    }
   \right)^{(1/2)\sqrt{3/2}}
                       \right\}    \nonumber  \\
  && = { {\tilde B} \over 2 \eta_0} \left\{
   \left( {       \sqrt{ 1 + (24/{\tilde B}^2) e^{8 \sigma} } + 1
            \over \sqrt{ 1 + (24/{\tilde B}^2) e^{8 \sigma} } - 1    }
   \right)^{(1/2)\sqrt{3/2}}
 - \left( {       \sqrt{ 1 + (24/{\tilde B}^2) e^{8 \sigma} } + 1
           \over \sqrt{ 1 + (24/{\tilde B}^2) e^{8 \sigma} } - 1    }
  \right)^{-(1/2)\sqrt{3/2}}  \right\}.
  \label{dilsolu}
 \eea
Finally from equation (\ref{axifirst}), we can find the solution
for axion
 \be
 \eta = {\eta_0 \over {\tilde B}} \left\{
   {     ( \sqrt{ 1 + (24/{\tilde B}^2) e^{8 \sigma} } + 1)^{\sqrt{3/2}}
       + ( \sqrt{ 1 + (24/{\tilde B}^2) e^{8 \sigma} } - 1)^{\sqrt{3/2}}
   \over ( \sqrt{ 1 + (24/{\tilde B}^2) e^{8 \sigma} } + 1)^{\sqrt{3/2}}
       - ( \sqrt{ 1 + (24/{\tilde B}^2) e^{8 \sigma} } - 1)^{\sqrt{3/2}}  }
  \right\}.
  \label{axisolu}
 \ee

\section{Brane cosmology}

In this section we will consider the cosmology probe D3-brane when
it is moving along a geodesic in the background type IIB solutions
of the previous section.

\subsection{ Without axion field }

The metric of D3-brane (\ref{ds10gen}) using the background
solution (\ref{5dmetwo}) is

 \be
   |g_{00}(r)| = e^{2\sigma} ,~~g(r) =  e^{2\sigma},
   ~~  g_{rr}(r) {L^2 d\sigma^2 \over 1 + (B^2 / 24) e^{-8\sigma}}
   ,~~g_S (r) = L^2 .
 \label{formofg}
 \ee
To apply the formalism of Sec. II we also need to express RR field
in terms of $\sigma$. From the ansatz for the RR field
 \be
 C_{0123} = C(r), ~~ F_{0123r} = {d C (r) \over d r} ,
 \ee
equation (\ref{RReq}) becomes
 \be
 {{d C(r)} \over dr} = 2 Q g^{2}g^{-5/2}_S \sqrt{g_{rr}}
 \ee
 where $Q$ is a constant.  Using the solution of the metric
 in (\ref{formofg}), the
$RR$ field can be integrated with appropriate normalization,
 \be
 C = \sqrt{ 1 + (24/B^2) e^{8 \sigma} }.
 \label{Csol}
 \ee

 Now we can calculate the effective density on the brane
 using equations (\ref{dilwosol}), (\ref{formofg}) and (\ref{Csol})
 \bea
 {8 \pi \over 3} \rho_{\rm eff}
  = {1 + ( 24/B^2 ) e^{-8 \sigma} \over L^2}
 && \Bigg[ \Bigg\{ E + \sqrt{1 + ( 24/B^2 ) e^{8 \sigma} } \Bigg\}^2
 e^{-8 \sigma}
 \Bigg( { \sqrt{1 + ( 24/B^2 ) e^{8 \sigma} } + 1
    \over \sqrt{1 + ( 24/B^2 ) e^{8 \sigma} } - 1  }
    \Bigg)^{\sqrt{3/2}}    \nonumber \\
 && - \bigg\{ 1 + {\ell^2 \over L^2} e^{-6 \sigma}
  \Bigg( { \sqrt{ 1 + (24/B^2) e^{8 \sigma }} + 1
    \over  \sqrt{1 + (24/B^2) e^{8 \sigma}}  -1   }
                   \Bigg)^{\sqrt{3/2}}  \Bigg\}  \Bigg].
 \eea
 If we rescale with $B^2=24$ and using $a=e^\sigma$, we get
 \bea
 {8\pi \over 3}\rho_{\rm eff} =
  {1 + a^{-8} \over L^2}
 && \Bigg[  \Bigg( \sqrt{1 + a^{-8} } + {E \over a^4}  \Bigg)^2
  \Bigg( { \sqrt{1 + a^{8}} + 1
    \over  \sqrt{1 + a^{8}}  -1  }
                   \Bigg)^{\sqrt{3/2}}  \nonumber \\
   && - \Bigg\{ 1 + {\ell^2 \over L^2} a^{-6}
  \Bigg( { \sqrt{1 + a^{8}} + 1
    \over  \sqrt{1 + a^{8 }}  -1  }
                   \Bigg)^{\sqrt{3/2}} \Bigg\} \Bigg],
 \label{rhoeff}
 \eea
 where the range of $a$ is $0 < a < \infty$, while the range of
 $\sigma$ is $- \infty < \sigma < \infty$.
When the universe brane is moving towards the singularity ($\sigma
\to - \infty, ~ a \to 0$) the universe is contracting while it is
moving outward ($\sigma \to \infty, ~a \to \infty$) it is
expanding. We also can calculate the scalar curvature of the
four-dimensional universe from (\ref{4dcurv}).

 Far from the black brane, one can see that $\rho_{\rm
eff} \sim a^{-4}$. The cosmological expansion due to the brane
motion is indistinguishable from the one by radiation on the
brane. This is the idea of the mirage cosmology.
 If we use the effective density
(\ref{rhoeff}) it blows up $\rho_{\rm eff} \sim a^{-8(2 +
\sqrt{3/2})}$ as $a \to 0$. Also if we move $a \to 0$, the
ten-dimensional metric becomes
 \be
 ds^{2}_{10}= a^2 (-dt^{2} + (d\vec{x})^{2})+
      {L^2 da^{2} \over a^2 } +  L^2 d\Omega_{5},
 \ee
which is an $AdS_5 \times S^{5}$ space. Thus the brane develops an
initial singularity as it reaches $a =0 $ where the description of
our formalism breaks down.

\subsection{ With axion field}

In this case, the only difference on the effective density comes
from the form of the dilaton. Still with ${\tilde B}^2 = 24,~
e^\sigma = a$, we have

\bea
  {8\pi \over 3}\rho_{\rm eff}
  && =
  {1 + a^{-8} \over L^2}    \nonumber   \\
 && \times \Bigg[ \Bigg( {12 \over \eta_0} \Bigg)^2
    \Bigg( \sqrt{1 + a^{-8} } + {E \over a^4}  \Bigg)^2
 \Bigg\{
 \Bigg( { \sqrt{1 + a^{8}} + 1
        \over  \sqrt{1 + a^{8}}  -1  }  \Bigg)^{\sqrt{3/2}}
+ \Bigg( { \sqrt{1 + a^{8}} + 1
        \over  \sqrt{1 + a^{8}}  -1  }  \Bigg)^{-\sqrt{3/2}}
        - 2 \Bigg\}        \nonumber \\
   &&~~~~ - 1 - {\ell^2 \over L^2} \Bigg( {12 \over \eta_0} \Bigg)^2 a^{-6}
\Bigg\{
 \Bigg( { \sqrt{1 + a^{8}} + 1
        \over  \sqrt{1 + a^{8}}  -1  }  \Bigg)^{\sqrt{3/2}}
+ \Bigg( { \sqrt{1 + a^{8}} + 1
        \over  \sqrt{1 + a^{8}}  -1  }  \Bigg)^{-\sqrt{3/2}}
        - 2 \Bigg\}      \Bigg].
 \label{rhoeffw}
 \eea
Near the brane the effective density blows up $\rho_{\rm eff} \sim
a^{-8(2 + \sqrt{3/2})}$ as $a \to 0$ which has the same functional
dependence as in the case without axion. However, far from the
brane ($a \to \infty$), it gives $\rho_{\rm eff} \sim - 1/L^2$.
This negative cosmological constant means that the expansion of
the universe stops at somewhere and eventually recollapses.
Comparing equation (\ref{rhoeffw}) with (\ref{rhoeff}), we see
that the effective density becomes negative faster if there is
axion field. The presence of the axion field do not play any
important role in the early stage of the evolution but its
coupling to other field gives different evolution at late stage.

\section{Discussion}

We considered the motion of a brane universe moving in a
background bulk space of type IIB string theory. For two different
backgrounds which give nontrivial dilaton profile, one without
axion field and the other with axion, we have derived the
Friedman-like equations. These give the cosmological evolution
which is similar to the one by matter density on the universe
brane. As the brane moves towards the singularity (smaller values
of radial coordinate) it contracts and while if it moves away from
the black brane it expands. So an observer on the three-brane will
see that the universe is expanding. The presence of axion field in
the background changes the dilaton profile but it does not change
the induced metric. Since dilaton, as well as the induced metric,
plays an important role in the effective density, the cosmological
evolutions are different for two different backgrounds. For both
cases, the effective density blows up as we move toward the
singularity showing the initial singularity problem and becomes
negative due to the angular momenta $\ell^2$ on the brane meaning
the recollapse of the universe. The functional dependence on the
radial coordinate shows that when there is axion field in the
ambient space the recollapsing of the universe occurs faster
compared with the case without axion field. It seems that this
phenomenon is true if we do the same calculation with field other
than axion.

\section*{Acknowledgement}

I would like to thank N. Kaloper, S. P. Kim and E. Kiritsis for
discussions and suggestions.


\begin{references}

\bibitem{RT}
T. Regge and C. Teitelboim, Marcel Grossman Meeting on General
Relativity, Trieste 1975, North Holland ;
 V. A. Rubakov and M. E. Shaposhnikov, Phys. Lett.
{\bf{B125}}, 136 (1983).

\bibitem{HDD}
N. Arkani-Hamed, S. Dimopoulos and G. Dvali, Phys. Lett.
{\bf{B429}}, 263 (1998), hep-ph/9803315; Phys. Rev. {\bf{D59}},
086004 (1999), hep-ph/9807344; I. Antoniadis, N. Arkani-Hamed, S.
Dimopoulos and G. Dvali, Phys. Lett. {\bf{B436}}, 257 (1998),
hep-ph/9804398.

\bibitem{RS}
L. Randall and R. Sundrum, Phys. Rev. Lett. {\bf{83}}, 3370
(1999), hep-ph/9905221; Phys. Rev. Lett. {\bf{83}}, 4690 (1999),
hep-th/9906064.

\bibitem{KVK}
N. Kaloper, Phys. Rev. {\bf{D60}}, 123506 (1999); hep-th/9905210;
N. Arkani-Hamed, S. Dimopoulos, G. Dvali, and N. Kaloper, Phys.
 Rev. Lett. {\bf 84}, 586 (2000), hep-th/9907209; H. B. Kim and H.
D. Kim, Phys. Rev. {\bf {D61}} 064003 (2000), hep-th/9909053; D.
N. Vollick, hep-th/9911181; S. Nam, hep-th/9911237; N. Kaloper,
Phys. Lett. {\bf {B474}}, 269 (2000), hep-th/9912125; N.
Arkani-Hamed, S. Dimopoulos, N. Kaloper, and R. Sundrum ,
hep-th/0001197; H. B. Kim, hep-th/0001209; N. Kaloper, J. M.
March-Russell, G. D. Starkman, and M. Trodden, hep-ph/0002001; S.
P. de Alvis, hep-th/0002174.

\bibitem{Pol}
J. Polchinski, Phys. Rev. Lett. {\bf{75}}, 4724 (1995),
hep-th/9510017; {\it String theory} (Cambridge University Press,
 1998).

\bibitem{HW}
P. Horava and E. Witten, Nucl. Phys. {\bf{B460}}, 506 (1996)
hep-th/9510209.

\bibitem{Tye}
P. Binetruy, C. Deffayet and D. Langlois, Nucl. Phys. {\bf B565},
269 (2000), hep-th/9905012; G. Dvali and S. H. H. Tye, Phys. Lett.
{\bf{B450}}, 72 (1999); hep-ph/9812483; E. E. Flanagan, S. H. H.
Tye and I. Wasserman, hep-ph/9909373.

\bibitem{CR}
H. A. Chamblin and H. S. Reall, Nucl. Phys. {\bf B562}, 133
(1999), hep-th/9903225; A. Chamblin, M. J. Perry and H. S. Reall,
J. High Energy Phys. {\bf{09}}, 014 (1999), hep-th/9908047.

\bibitem{Kiritsis}
  A. Kehagias and E. Kiritsis, J. High
Energy Phys. {\bf{9911}}, 022 (1999), hep-th/9910174.

\bibitem{PP}
  E. Papantonopoulis and I. Pappa, hep-th/0001183.

\bibitem{GSW}
  M. B. Green, J. H. Schwarz, and E. Witten,
  {\it String Theory}, Vol. 2
  (Cambridge University Press, Cambridge, England, 1987).

\bibitem{GKT}
  S.S. Gubser, I. R. Klebanov, and A. A. Tseytlin,
  Nucl. Phys. {\bf B534}, 202 (1998), hep-th/9805165.

\bibitem{GGP} G. W. Gibbons, M. B. Green and  M. J. Perry,
Phys. Lett. {\bf B370}, 37 (1996), hep-th/9511080.

\bibitem{Gub}
A. Kehagias and K. Sfetsos, Phys. Lett. {\bf B454}, 270 (1999),
hep-th/9902125; S. S. Gubser, hep-th/9902155.

\bibitem{Kesf}
 A. Kehagias, K. Sfetsos, Phys. Lett. {\bf B456}, 22 (1999),
  hep-th/9903109


\end{references}
\end{document}